**Rejuvenation versus overaging: The effect of cyclic loading/unloading on the segmental dynamics of PMMA glasses**

**Trevor Bennin*, Enran Xing, Josh Ricci, M.D. Ediger**

**Department of Chemistry, University of Wisconsin – Madison, Madison, Wisconsin 53706, United States**

***Corresponding Author (tbennin@wisc.edu)**

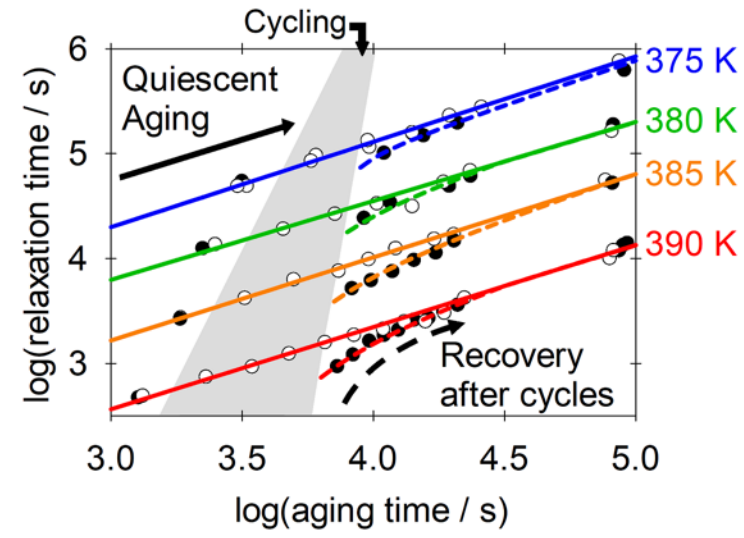

**For Table of Contents use only**

## Abstract

The acceleration of structural relaxation or physical aging by deformation, known as overaging, has been reported in experiments and simulations of polymer and colloid glasses, and correctly accounting for overaging is important for the prediction of the long-term behavior of polymer glasses in engineering applications. Here the effects of cyclic loading/unloading on the segmental dynamics and mechanical properties of PMMA glasses are investigated using a probe reorientation technique and time-aging time superposition of the mechanical response, respectively. Sets of 5000 tensile loading/unloading cycles were performed at temperatures between $T_g$ – 10 K and $T_g$ – 25 K with cycle extension strains ranging from 0.003 to 0.007. After cycling, the segmental dynamics measured with the probe reorientation technique either remained unchanged or were faster relative to an undeformed sample. The relaxation times of cycled glasses recovered with a common time scale on the order of their aging time, indicating that they retain a memory of their original age, as opposed to a full erasure of their thermal and mechanical history. Surprisingly, changes as a result of cycling were more obvious in probe reorientation measurements than in the mechanical properties, suggesting that the probe reorientation technique can sensitively detect nonlinear effects of deformation. No evidence of overaging was observed in the optical or mechanical measurements as a result of these cyclic loading/unloading experiments.

## Introduction

Polymer glasses are commonly utilized for their light weight and mechanical toughness. However, the properties of polymer glasses evolve over time due to their nonequilibrium nature, a feature common to all glasses. This slow evolution, known as structural recovery or physical aging, can ultimately affect the use of polymer glasses in modern technologies by changing, for example, their gas permeability[1, 2] or dielectric strength.[3-6] Physical aging also causes stiffening[7] and embrittlement[8, 9] in polymer glasses, which can lead to catastrophic failure in load-bearing applications. To minimize failure of polymer glass materials, it is important to prevent or predict the changes that occur during long-term use.

A significant thrust of recent research has focused on manipulating the ductility of solid polymers to enable their use in advanced engineering applications. Adjusting crystallinity,[10-12] loading with nanoparticles,[13, 14] and blending with plasticizers[15-17] have all been shown to enhance the ductility of polymer glasses. In addition, pre-deformation of a polymer in the melt state[9, 18] or the glassy state[8] can also make a polymer glass more ductile. In the glassy state, this deformation-induced ductility is often described as “rejuvenation”, i.e., the deformation apparently reverses the effect of aging and prepares a glass state that appears to be less annealed (although other interpretations have been advanced[19]).

Conceptually, rejuvenation acts to pull a system up the potential energy landscape to a position typically occupied by a less aged system, where it may exhibit greater ductility, higher enthalpy and faster dynamics.

Conversely, there is a growing number of simulations[20-28] and experiments[29-35] indicating that mechanical deformation can cause a glass to show properties of a more annealed or aged glass. In simulated glasses, certain types of deformation can cause the glass to have a lower energy than an undeformed glass would achieve from physical aging over the same time period.[20-28] This accelerated aging process is known as overaging. A simple explanation for these results is that deformation can enhance molecular mobility in glasses, and the resulting enhanced mobility may then allow for the molecular motions associated with physical aging to occur more quickly.

There is strong evidence that overaging can occur in glassy colloids.[29] In these experiments, microscopic particle rearrangements were measured directly with multispeckle diffusing wave spectroscopy. It was found that the particle rearrangements within deformed systems, measured after the deformation has ceased, could be slower than those of a quiescent system of the same age. This implies that the deformed colloidal glass reached a larger effective age, consistent with overaging.

A number of experiments on polymer glasses have observed unusual mechanical properties in deformed samples that have been interpreted as evidence for overaging. In these experiments, deformations can cause longer stress relaxation rates[32], lower compliances[33] or larger yield stresses[30, 34, 35] than a similarly aged quiescent glass. Because these changes are typically associated with aging, the modified properties are interpreted as evidence of overaging, or mechanically-enhanced aging[32, 33] or stress-accelerated aging.[35] However, since the results of nonlinear mechanical experiments can often be interpreted in multiple ways,[36] these results do not constitute unambiguous evidence of overaging. For example, the observation of a longer relaxation time scale for a linear stress relaxation experiment would directly indicate slower (and more aged) segmental dynamics,[37] but a similar conclusion cannot be drawn for nonlinear deformation.[38] Similarly, an increased yield stress can result from aging but it can also occur as a result of chain orientation.[39] It is important to understand when and how overaging can occur. For polymer glasses, the effects of overaging have been incorporated into a constitutive model used to estimate the time-to-failure under static or oscillating stress,[35, 40] suggesting an important role for overaging in the failure in these systems. In addition, the ability to create a more annealed glass through deformation may be useful in creating thermodynamically stable bulk glasses.[21]

Here we attempt to understand overaging in polymer glasses in a new way by investigating the changes in segmental dynamics brought about by mechanical deformation. For many types of glasses, material properties and the time scale for physical aging are thought to be intimately related to the alpha relaxation process (which is determined by segmental motions in polymer glasses).[7] Given that segmental motions allow aging to occur, it is reasonable to expect that experimental access to the segmental dynamics of a polymer glass could provide fundamental insight into the overaging process. Segmental dynamics are measured directly using a probe reorientation technique; previous work using this technique has quantified changes in the segmental dynamics of a polymer glass due to physical aging[41, 42] and during active deformation,[43-46] providing direct access to the microscopic dynamics of the glass. In our investigation of overaging, we look for deformation that slows dynamics more rapidly than the slowing that would have occurred by physical aging in the absence of deformation. To plan these experiments, we identified some of the important features of computer simulations that lead to clear overaging: rapid quenching[20, 22-28] to low deformation temperatures;[20-28] use of small,[20, 21, 23, 26, 27] symmetrical cycles;[21-23, 25-28] and total cycling times that exceed the age of the system before cycling occurs.[20, 22, 25, 26]

In this work, we investigated the effects of cyclic loading/unloading on the segmental dynamics of polymer glasses. We performed tensile loading/unloading cycles at 1 Hz with cycle extension strains ranging from 0.003 to 0.0072 on lightly crosslinked PMMA glasses at temperatures ranging from $T_g$ – 10 K to $T_g$ – 25 K. We measured changes in the segmental dynamics and mechanical properties of the cycled glasses using the probe reorientation technique and time-aging time superposition of the mechanical response, respectively. We find that small amplitude asymmetric cycling causes the segmental dynamics of PMMA glasses to speed up or stay the same relative to an aging sample but does not cause overaging. In addition, we find that the recovery of relaxation times of cycled PMMA glasses indicates that they retain memory of their original age, as opposed to a full erasure of their thermal and mechanical history. Surprisingly, the probe reorientation technique measured clear changes in the relaxation time under conditions where the changes to the mechanical properties were more consistent with linear behavior. This suggests that the probe reorientation technique can sensitively detect nonlinear effects of deformation.

## Experimental

**Sample Preparation.** Lightly crosslinked PMMA samples were prepared using free radical polymerization, as previously published.[41] Briefly, methyl methacrylate, ethylene glycol dimethacrylate and the optical probe N,N'-dipentyl-3,4,9,10-perylenedicarboximide were combined in solution. Then, benzoyl peroxide, an initiating agent, was added, and the solution was partially polymerized at 343 K for 30 minutes. The partially polymerized solution was pressed between two microscope slides and transferred to a vacuum oven, where the polymerization process continued for 24 hours at 343 K and then another 24 hours at 393 K, all under nitrogen. After polymerization had completed, the polymer films were removed from the microscope slides by sonication. "Dog bone" shaped samples were cut from these films using a custom-made steel die that conforms to a scaled-down version of ASTM D1708-10. Samples used in this work had an inhomogeneous thickness profile; thicknesses ranged from 20 to 30 $\mu$m at the thinnest point, and the difference between the thinnest and thickest regions of a given sample was 10 $\mu$m at most. The glass transition temperature of the samples was measured to be $\mathrm{T_g} = 400 \pm 1$ K as determined by DSC, using the midpoint of the transition of the second heating scan (with heating and cooling performed at 10 K/min).

**Thermal Protocol.** The PMMA sample was clamped into clips and loaded into the temperature-controlled cell of the deformation apparatus, where it was kept during all stages of the experiment. Samples were annealed at 415 K for 30 minutes to erase thermal and mechanical history. After annealing, samples were cooled at 1 K/min to the testing temperature and then held isothermally for the remainder of the experiment. The aging time of the sample, $t_a$, was determined from the time that had passed after the temperature of the sample crossed $T_g$. The temperature was stable to $\pm 0.2$ K once the testing temperature was reached. Melting point tests indicate that the temperature accuracy of the cell is $\pm 1$ K.

**Probe Reorientation Measurements.** The segmental dynamics of PMMA glasses before and after cycling were tracked using the probe reorientation technique, which has been described previously.[43] Briefly, fluorescent probes (DPPC) are dispersed in the PMMA sample at a concentration of about $10^{-6}$ M with an isotropic distribution of orientations. The sample is exposed to a linearly polarized laser beam that preferentially photobleaches probes with transition dipoles that are aligned to the polarization state of the beam, creating an anisotropic distribution of orientations in the probes. Then, fluorescence of the remaining unbleached probes is induced by a weak, circularly polarized beam. The fluorescence intensities parallel and perpendicular to the polarization state of the photobleaching beam are then used to calculate the anisotropy of the orientation distribution of the probes as a function of time, $r(t)$.

The relaxation of the anisotropy can be described by the Kohlrausch-Williams-Watts (KWW) function:

$$r(t) = r(0)e^{-\left(t/\tau\right)^{\beta}}$$

Here, $r(0)$ is the initial anisotropy induced by photobleaching, $\tau$ is a characteristic probe reorientation relaxation time, and the $\beta$ parameter characterizes the nonexponentiality of the decay, which is related to the width of the distribution of relaxation times of the system.

In this work, we report the relaxation time, $\tau$, extracted from KWW fits to the anisotropy data. It has been reported previously that reorientation of DPPC is a good reporter of segmental dynamics in lightly crosslinked PMMA in the glassy state.[41] Prior work using the probe reorientation technique during active deformation also reports $\beta$ to describe changes to the underlying distribution of relaxation times in the glass.[44-46] In this work, unconstrained fits using the KWW function gave $\beta$ values very near each other (within 0.04) and slightly less than $0.31$, the equilibrium value obtained from previous probe reorientation measurements in lightly crosslinked PMMA.[44] To minimize the uncertainty in fitted relaxation times, we constrained $\beta$ = $0.31$ for the results reported here.

All probe reorientation measurements are performed at the aging temperature in the temperature-controlled deformation apparatus to directly measure segmental dynamics before and after mechanical cycling. The time required to measure a single anisotropy decay function increases with decreasing temperature and increasing aging time, and ranges from 10 min to 150 min.

**Mechanical Protocol.** Computer control of the linear actuator used in the deformation apparatus allows for cyclic loading/unloading of the PMMA glasses. The cyclic loading/unloading was performed in three steps. First the sample was extended by a fixed engineering strain, $\varepsilon_{cycle}$, at a constant strain rate. Second, the sample was retracted at the same rate until the sample reached zero stress, as measured by a 5 N load cell. Finally, all motion was temporarily halted (for about 0.1 s), and the position of the linear actuator was recorded. Cyclic loading/unloading of the sample was performed at a constant engineering strain rate $\dot{\varepsilon}_{cycle} = 7.67 \times 10^{-3}, 1.53 \times 10^{-2}$ or $1.84 \times 10^{-2}$ $s^{-1}$ for $\varepsilon_{cycle} = 0.003, 0.006$ or $0.0072$, respectively. Together, the three steps of cycling took 1 second, leading to an effective frequency $f = 1$ Hz. These three steps were repeated 5000 times for one complete set of cycles. A representative example of a subset of 5000 cycles are shown in Figure 1. For reference, we note that the yield strain for the samples used in these experiments is $\varepsilon_{yield} = 0.03$ for constant strain rate extension at $\dot{\varepsilon} \approx 3 \times 10^{-5}$ $s^{-1}$, indicating that the values of $\varepsilon_{cycle}$ range from $0.10\varepsilon_{yield}$ to $0.24\varepsilon_{yield}$.

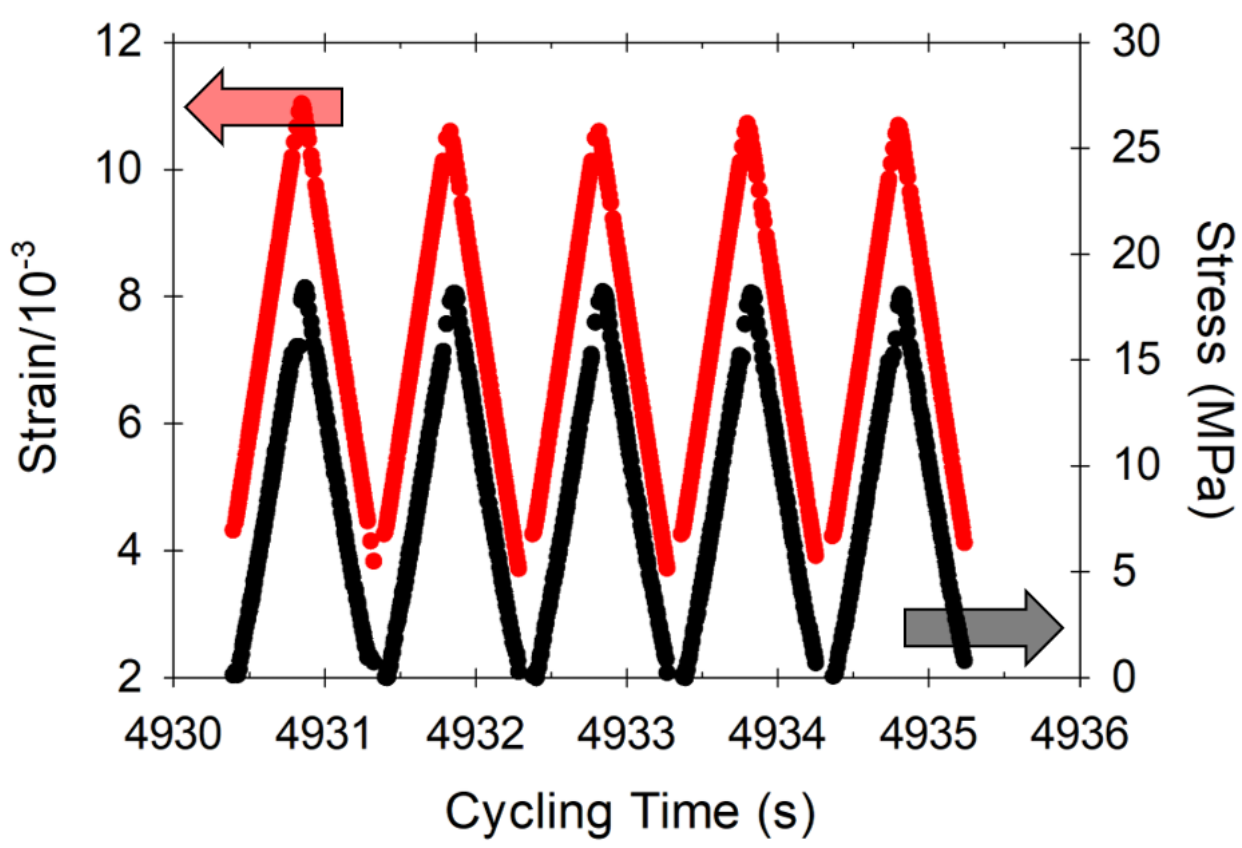

**Figure 1.** Stress and strain data from a subset of 5000 tensile loading/unloading cycles on PMMA at 375 K.

For the results presented in this work, two cycling protocols were used. The first protocol was to subject the sample to one set of 5000 cycles after the initial probe reorientation measurement. Following this single set of cycles, a series of probe reorientation measurements tracked the time evolution of the segmental dynamics of the sample. The second protocol was to subject the sample to 3 sets of 5000 cycles after the initial probe reorientation experiment. Following each set of cycles, a single probe reorientation measurement was performed, after which the next set of cycles was initiated. This second protocol allowed a comparison between the segmental dynamics of the sample (using the probe reorientation technique) and the mechanical response at multiple points throughout the experiment.

We present the mechanical data as the ratcheting compliance. (Ratcheting is the buildup of anelastic strain with asymmetric cycling deformation.[47]) The ratcheting strain is defined as $\varepsilon_{rat} = \frac{1}{2}(\varepsilon_{max} - \varepsilon_{min})$, which is equivalent to $\varepsilon_{rat} = \varepsilon_{min} + \frac{1}{2}\varepsilon_{cycle}$ for a fixed forward extension of $\varepsilon_{cycle}$. The ratcheting compliance is calculated as $J_{rat} = \frac{\varepsilon_{rat}}{\sigma_{average}}$, where $\sigma_{average}$ is the average stress over the entire set of cycles. We have analyzed the ratcheting compliance using the time-aging time superposition principle. In the linear response regime, this analysis provides a method to assess changes in the segmental dynamics during aging. This analysis is commonly used on creep compliance data from a small constant stress deformation (within the linear regime). In the Supplemental Information, we show that the ratcheting compliance is the same as the standard linear creep compliance for lightly crosslinked PMMA glasses at 375 K (after multiplication by a factor of 1.76).

Temperature rise due to energy dissipation in poly(methyl methacrylate) glasses during cyclic loading/unloading has previously been reported.[48] However, the temperature rise associated with the deformation in this work is expected to be negligible. Abar *et al.*[48] observed the temperature in their PMMA sample to increase from room temperature by about 6 K when the sample was cycled with a maximum load of 44.4 MPa at 0.5 Hz. Compared to the work by Abar *et al.*, the cycling done in this work is performed with lower loads (which will dissipate less energy) and a similarly low frequency, so the temperature rise is expected to be negligible here. In addition, the samples used in this work are approximately 400 times thinner than those from Abar *et al.* so heat dissipation should occur much more rapidly, especially in comparison to the relaxation time recovery time scales presented below.

## Results

**Single Set of 5000 Cycles.** Probe reorientation measurements were performed on PMMA glasses before and after being subjected to one set of 5000 tensile loading/unloading cycles. In Figure 2, it is seen that relaxation times of samples after cycling are shorter than undeformed samples and show no evidence of overaging. For the experiments shown in this figure, 5000 tensile loading/unloading cycles were performed with $\varepsilon_{cycle} = 0.006$ and $f = 1$ Hz. The relaxation time for cycled samples are the same as those for aging samples before the cycling occurs, which is expected. However, the relaxation times for cycled samples are decreased from the generic aging trajectory immediately after cycling for all temperatures. The decreased relaxation time is commonly called mechanical rejuvenation (though the precise nature of rejuvenation is a matter of debate,[19] we use the term only to denote an apparently reduced relaxation time in response to a mechanical perturbation). In addition, the amount by which relaxation times are decreased from the generic aging trajectory decreases with temperature. After the initial decrease, the relaxation times of the cycled samples recover to, but never increase above, those of the purely aging system, within experimental error. Thus, under these experimental conditions, no overaging was observed.

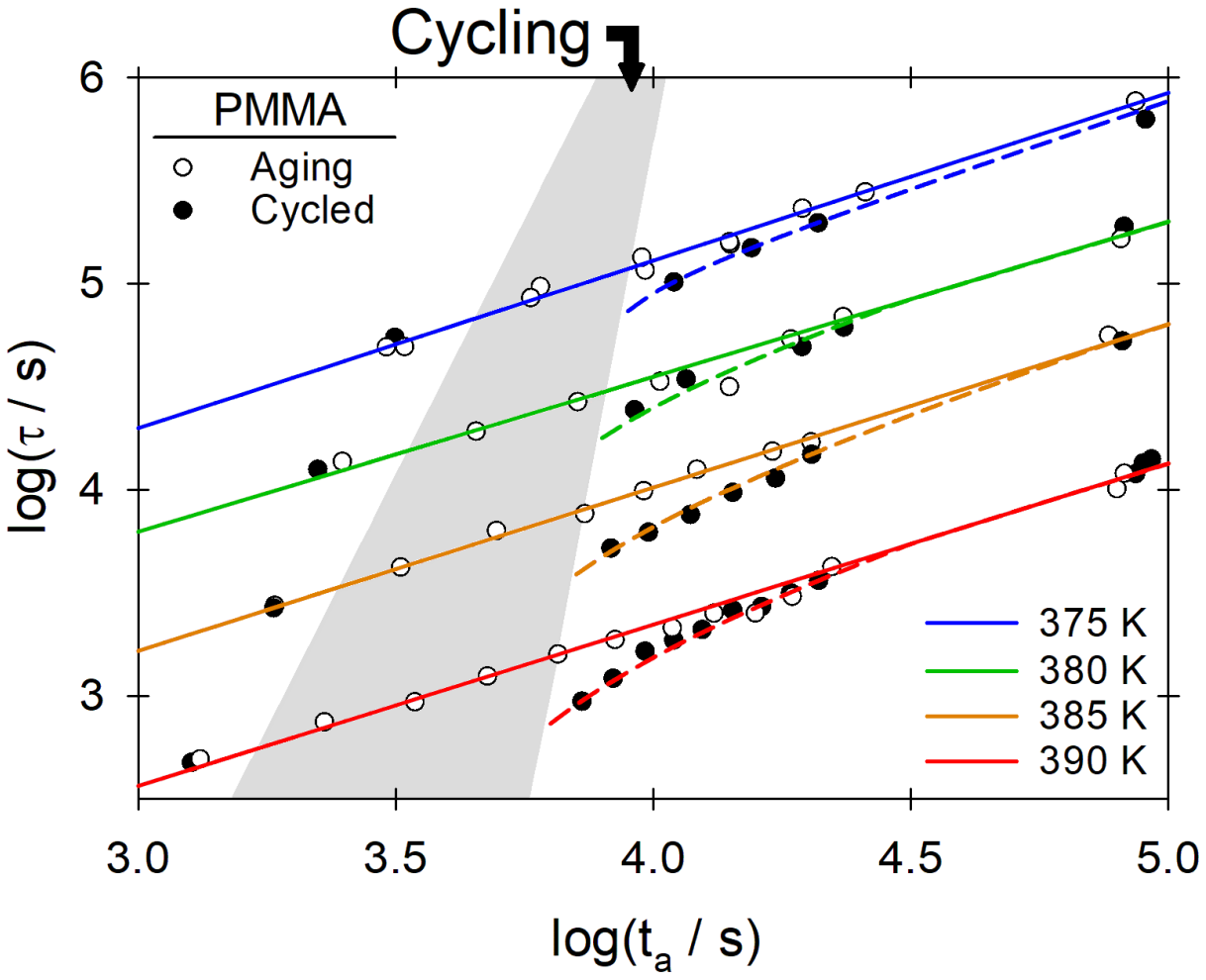

**Figure 2.** Evolution of the segmental relaxation time of PMMA glasses subjected to one set of 5000 loading/unloading cycles (shaded region), with $\varepsilon_{cycle} = 0.006$ and $f = 1$ Hz at temperatures ranging between $T_g$ – 10 K and $T_g$ - 25 K. Open symbols are collected from aging samples and closed symbols are collected from samples after cycling. Cycled glasses show decreased relaxation times immediately after cycling is ceased, then recovery toward the generic aging trajectory, showing no signs of overaging. The dashed lines are guides to the eye, representing the recovery of relaxation times.

**Multiple Sets of 5000 Cycles.** Experiments were also performed in which PMMA glasses were subjected to three sets of 5000 tensile loading/unloading cycles. Probe reorientation measurements were performed before and after each of the three sets of 5000 cycles. Figure 3 shows the relaxation times of these samples. The loading/unloading cycles for these measurements were performed with the same parameters as the single set of cycles shown in Figure 2 ($\varepsilon_{cycle} = 0.006$ and $f = 1$ Hz). The only difference is that these samples were subjected to three sets of cycles, rather than just one, to maximize the cycling time during a single experiment. For each temperature, it is seen that the sequential loading/unloading cycles tends to decrease relaxation times incrementally further from the pure aging trajectory. After one set of cycles, the extent of rejuvenation in the cycled samples of Figure 3 is consistent with that of the cycled samples in Figure 2, which is expected at this stage of the experiment where the thermal and mechanical history were the same. For both a single set of cycles and multiple sets of cycles, the smallest amount of rejuvenation occurs at the coldest experimental temperatures. At no point do the relaxation times of cycled samples become longer than the purely aging samples as would be expected for an overaged sample.

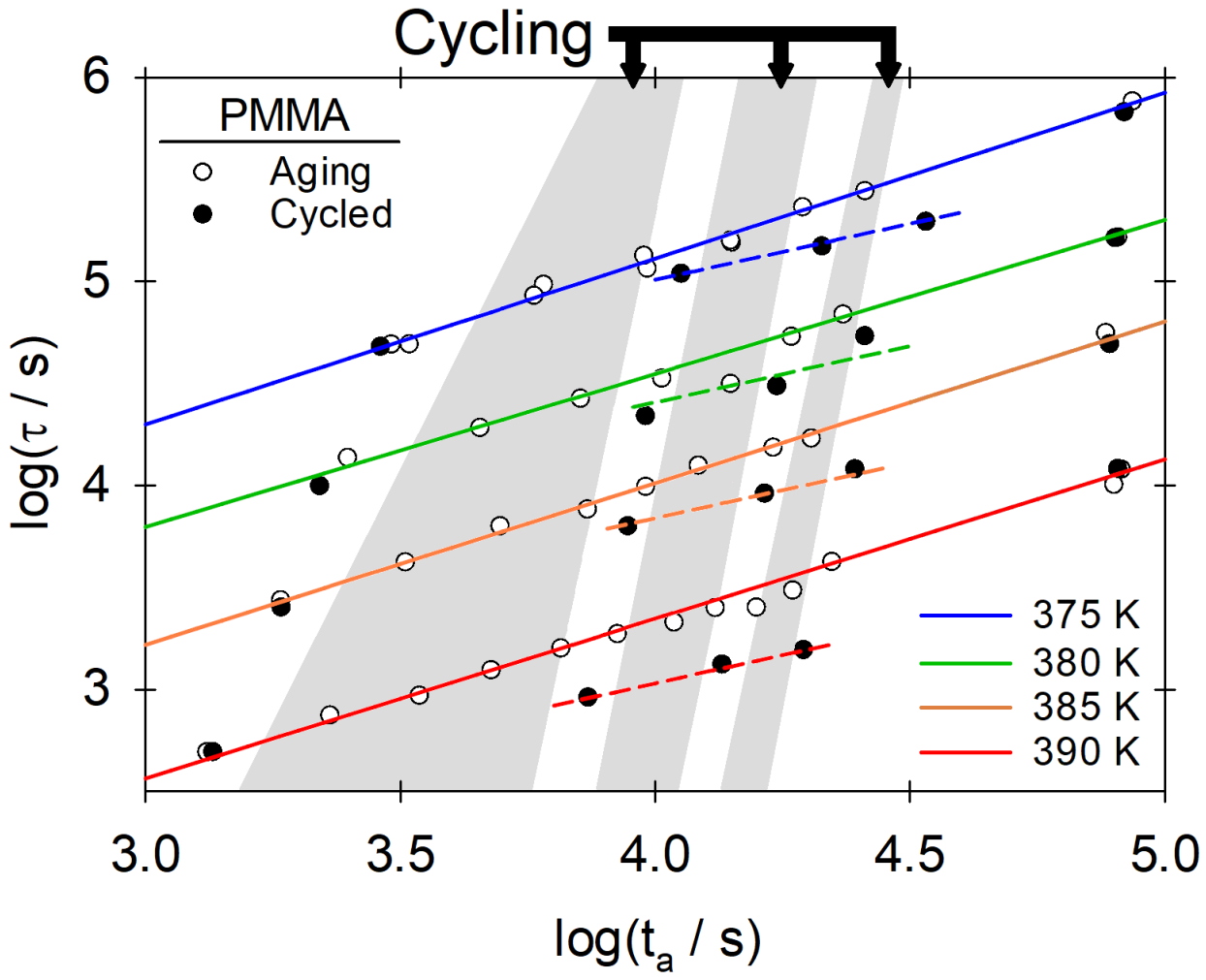

**Figure 3.** Evolution of the segmental relaxation time of PMMA glasses subjected to three sets of 5000 loading/unloading cycles (shaded regions) with $\varepsilon_{cycle} = 0.006$ and $f = 1$ Hz. Relaxation times from cycled samples are incrementally decreased from the aging trajectory with each subsequent set of cycles, indicating that the cycling does not induce overaging. The dashed lines are guides to the eye.

In the literature, overaging is most strongly identified at cold temperatures near some critical strain amplitude.[21, 23, 26] To investigate the strain amplitude dependence, we have performed measurements with various $\varepsilon_{cycle}$ at our lowest experimental temperature. Figure 4 shows that relaxation times of PMMA glasses after cyclic loading/unloading with various $\varepsilon_{cycle}$ at 375 K are either unchanged or decreased from the undeformed aging sample. Again, samples were subjected to three sets of 5000 cycles to maximize the cycling time during an experiment. For all $\varepsilon_{cycle}$ explored, the measured relaxation times are the same as or less than the undeformed aging sample. The extent of rejuvenation relative to the aging sample decreases with $\varepsilon_{cycle}$, until there is no discernible rejuvenation with the lowest amplitude cycles explored.

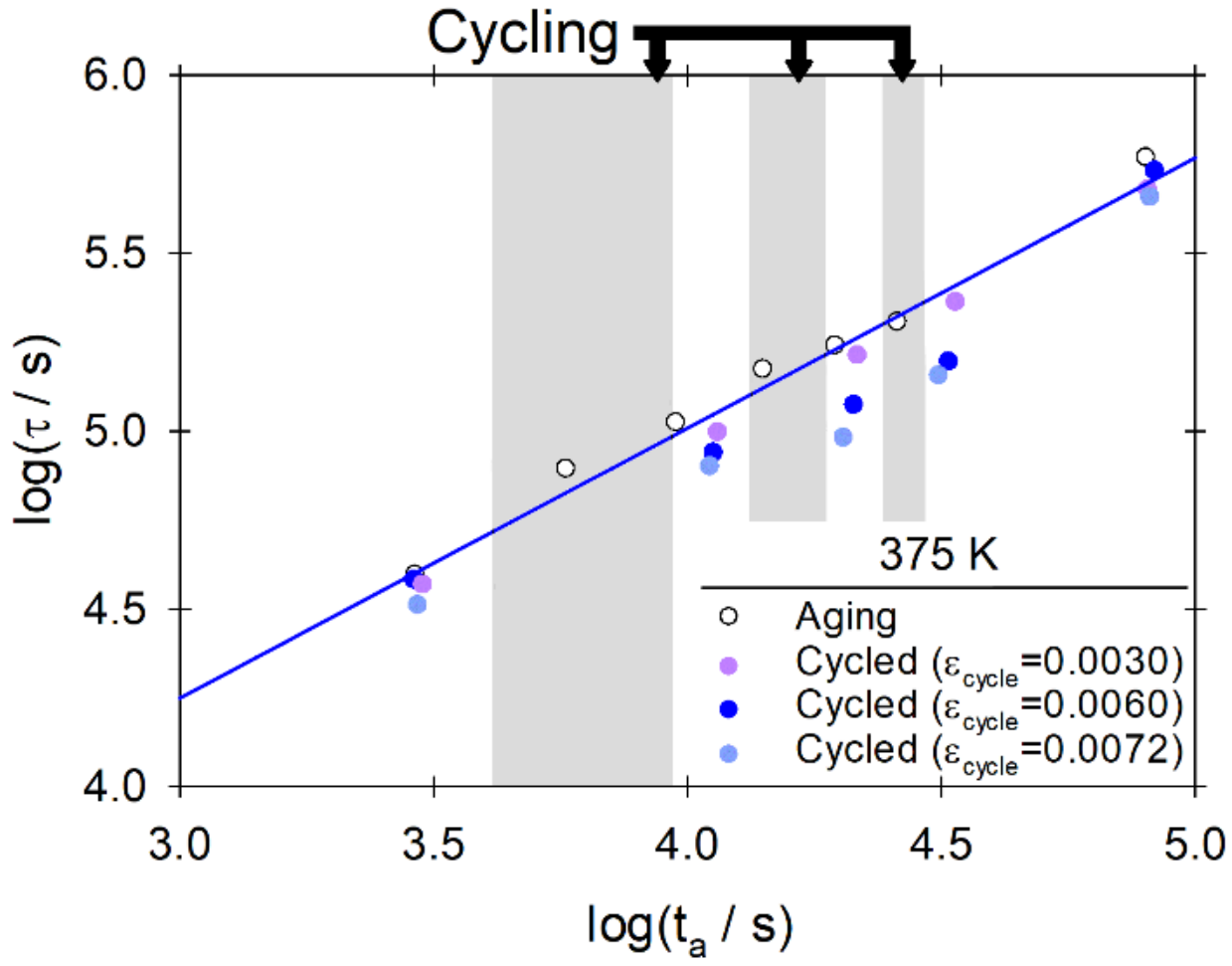

**Figure 4.** Evolution of the segmental relaxation time of PMMA glasses subjected to three sets of 5000 loading/unloading cycles (shaded regions) with various $\varepsilon_{cycle}$ at 375 K. The deviation of the measured relaxation times varies with $\varepsilon_{cycle}$, but no value of $\varepsilon_{cycle}$ induces longer relaxation times than the undeformed sample.

In addition to probe reorientation data, the cyclic loading/unloading procedure provides access to the ratcheting creep compliance (calculated from mechanical data in the shaded regions of Figure 3). The evolution of the ratcheting compliance due to combined effects of physical aging and cyclic loading/unloading can be investigated using the time-aging time superposition principle. As recommended by Struik for long-term mechanical tests,[7] we limited the mechanical response used for superposition to the momentary creep response, where $t_{creep}/t_a \leq 0.1$, to minimize the impact of aging that occurs during the deformation.

The ratcheting compliance shown in Figure 5 reasonably obeys time-aging time superposition. For these superpositions, both horizontal and vertical shifts were employed. It has been argued that there is no physical justification for vertical shifts in time-aging time superposition;[49] however, the vertical shifts employed here were small and are attributed to measurement error for all temperatures except 390 K (see the Supplemental Information for further details). Measurement error is magnified at the earliest creep times and coldest temperatures since the anelastic part of the ratcheting deformation is only a few microns, as exemplified by the green curve of the 375 K data in Figure 5. Beyond this limit, the curves are readily superimposable.

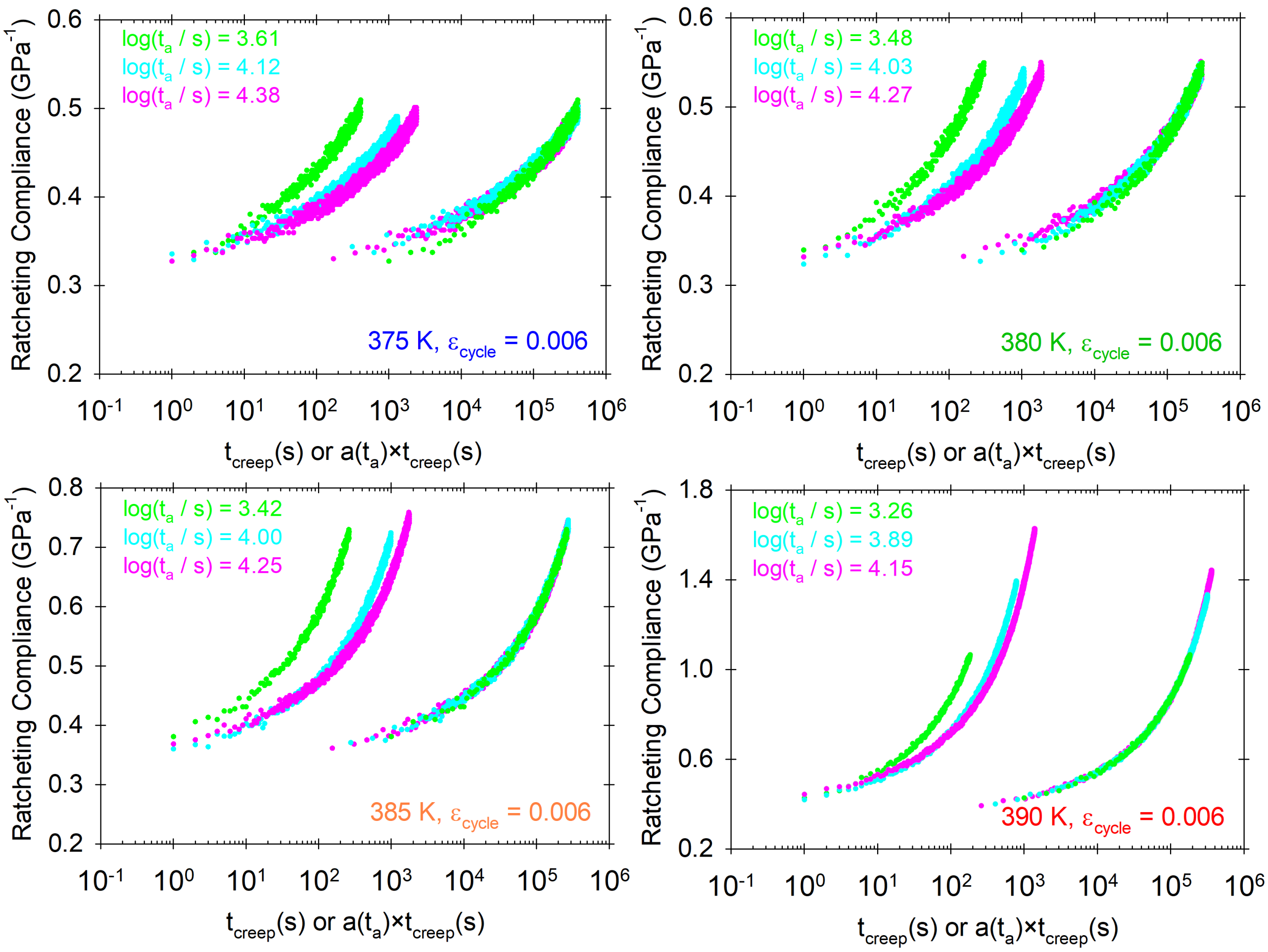

**Figure 5.** Ratcheting compliance of PMMA glasses subjected to three sets of cyclic loading/unloading (same experiment as shown in Figure 3). For each temperature, the three curves shown utilize mechanical data collected from the beginning of each of the three cycles of deformation. Subsequent compliance curves evolve more slowly. At each temperature, the best superposition of the compliance curves is offset to the right for clarity.

The evolution of the time-aging time shift factors obtained from the superposition of the ratcheting compliance for PMMA glasses is shown in Figure 6. As we show below, the shift factors are consistent with a linear deformation and inconsistent with either rejuvenation or overaging. The aging rate of the ratcheting compliance is defined as $\mu = -\frac{d\log(a(t_a))}{d\log(t_a)}$ and is found to be between 0 and 1 for typical polymer glasses.[7] Figure 6 shows that the aging rate for the experiments shown in Figure 5 is nearly unity for all temperatures for deformation temperatures less than 390 K. In general, the aging rate of a polymer is 0 at $T_g$, as the sample can readily equilibrate on short time scales; then, as the temperature decreases, the aging rate increases towards 1, remains near 1 for some temperature range, and then decreases back towards 0.[7] The aging rates calculated using the ratcheting compliance are the same as those calculated using the standard linear creep compliance as shown below in Figure 9 (a description of how the standard creep compliance was measured can be found in the Supplemental Information). Thus, the changes in the ratcheting compliance are consistent with a linear deformation and inconsistent with either rejuvenation, which is expected to lower the observed aging rate, or overaging, which is expected to increase the observed aging rate.

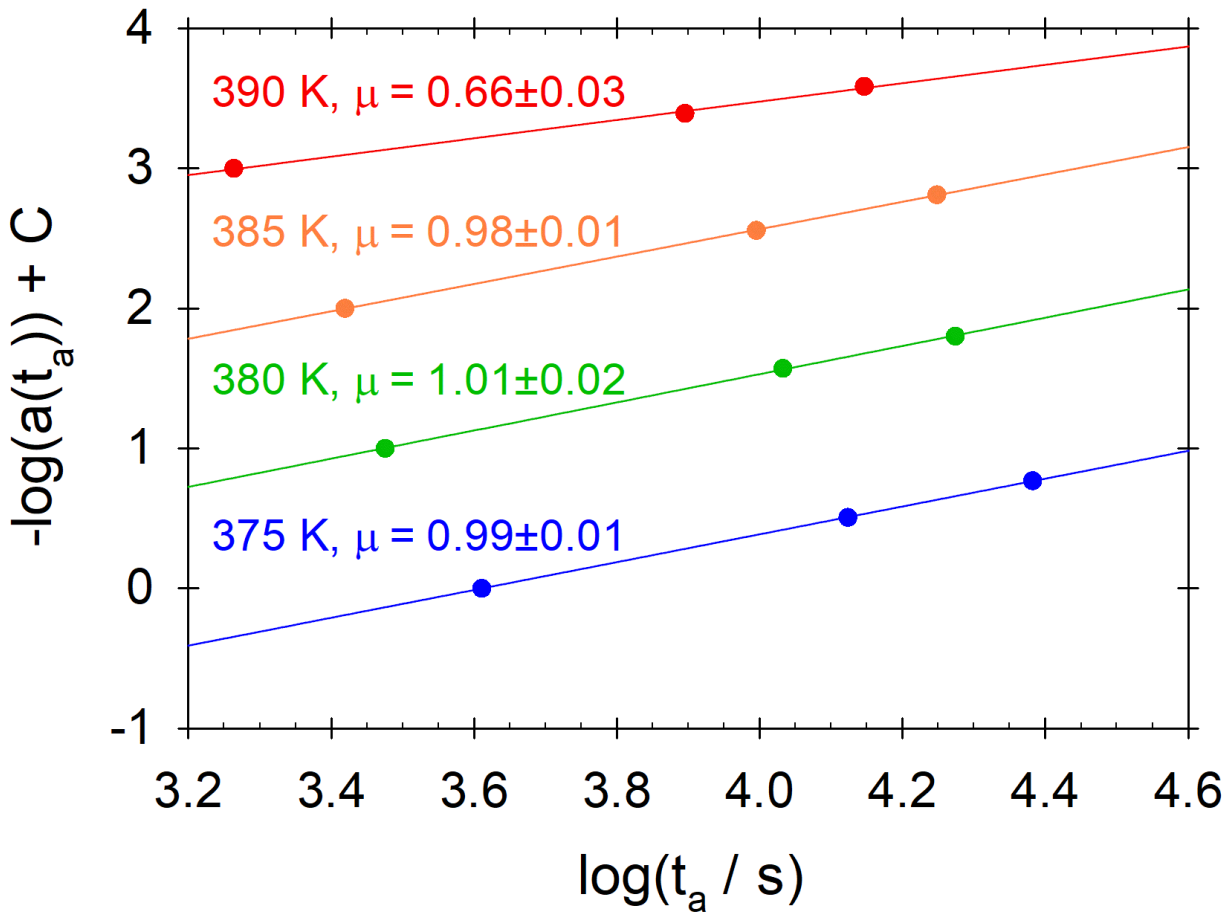

**Figure 6.** Time-aging time shift factors for ratcheting compliance superposition in PMMA glasses at various temperatures with $\varepsilon_{cycle} = 0.006$. The data points are vertically offset for clarity. The aging rate, $\mu$, is the slope of a linear fit through the data points. The uncertainty represents the standard error of the slope from the fitting procedure.

The time-aging time superposition process can also be used to analyze the ratcheting compliance from various $\varepsilon_{cycle}$ (calculated from mechanical data in the shaded regions of Figure 4). The aging rates from this superposition, seen in Figure 7, also do not show indications of overaging. The ratcheting compliance curves for cycles of various $\varepsilon_{cycle}$ were superimposed using the same procedure described earlier; the superpositions themselves can be viewed in the Supplemental Information. The aging rates for the investigated $\varepsilon_{cycle}$ values are the same, within error. This is consistent with the cycles being in the linear response regime. The observation of a constant aging rate as a function of $\varepsilon_{cycle}$ is inconsistent with either rejuvenation or overaging.

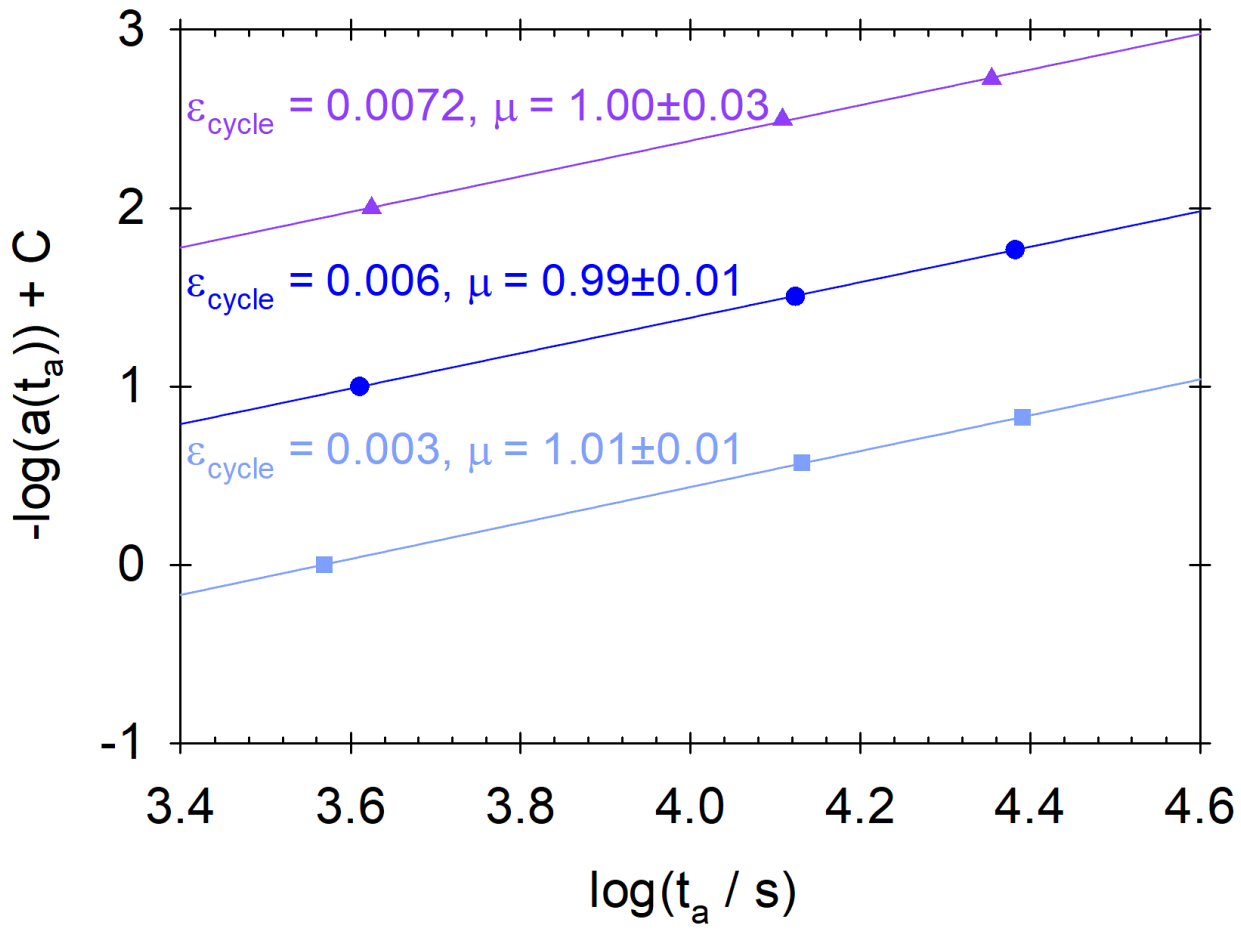

**Figure 7.** Time-aging time shift factors for ratcheting compliance superposition in PMMA glasses at 375 K with various $\varepsilon_{cycle}$. The data points are vertically offset for clarity. $\mu$ and its uncertainty are calculated as in Figure 6. The identical aging rates for the various $\varepsilon_{cycle}$ are consistent with these deformations being in the linear response regime.

## Discussion

The results shown here highlight the effects of cyclic loading/unloading on the segmental dynamics of PMMA glasses. While we could not explore the vast parameter space relevant for overaging, we did explore a range of temperatures just below the glass transition temperature, which is one relevant regime for polymer glasses. The probe reorientation data and the compliance data indicate that segmental dynamics and mechanical properties remain unchanged or are rejuvenated from the generic aging behavior. Relative to quiescent aging, we did not observe slower relaxation times or stiffer glasses. In other words, we do not find any evidence of overaging in these experiments.

It is interesting to note that the ratcheting compliance data in Figures 6 and 7 seems to be less sensitive to rejuvenation effects which are observed clearly and reproducibly in the probe reorientation data of Figures 2, 3, and 4. While probe reorientation showed that the relaxation times of cycled glasses were decreased from those of an undeformed, aging sample, which is indicative of nonlinear deformation, the aging rate of the ratcheting compliance was seemingly unaffected by the cycling, which is consistent with a linear deformation. However, neither the aging of the ratcheting compliance nor the probe reorientation data indicates the presence of any overaging effects under the investigated experimental conditions.

The lack of clear rejuvenation effects in the aging rate of the ratcheting compliance suggests that the probe reorientation technique is more sensitive to nonlinear deformation than the ratcheting compliance data collected during cycling. Prior work with the probe reorientation technique has similarly concluded that mechanical experiments are less sensitive to nonlinear deformation. Hebert and Ediger[50] measured the extent of rejuvenation in a PMMA glass after a reversing strain deformation using probe reorientation and the yield stress of a subsequent constant strain rate deformation. They observed that probe reorientation showed signs of rejuvenation (evidenced by decreased relaxation times relative to an undeformed sample) at pre-yield strain values for which the mechanical data showed no change in the yield stress.

In the rest of the discussion below, we will compare the results obtained here with those from other experiments and simulations.

**Rejuvenation and Recovery of Relaxation Times after Cycling.** Figure 8 shows that the segmental dynamics of PMMA glasses after cycling recover to the generic aging trajectory with a common time scale without crossing into the overaging regime. The format of Figure 8 is motivated by Smessaert and Rottler.[51] According to ref [51], there are four important time scales for understanding the evolution of the segmental dynamics of glasses after a mechanical perturbation: the waiting time, $t_w$ (the time span between when the glass is formed to the beginning of deformation); the deformation time, $t_d$; the total aging time before recovery, $t_a = t_w + t_d$; and, the recovery time, $t_r$. The x-axis of Figure 8 is a normalized recovery time, $t_r/t_a$, and the y-axis is the measured relaxation time at $t_r + t_a$ normalized by the expected relaxation time at $t_a$ (approximated from a fit to the aging data). The three colored regions of the plot are defined by their boundaries. The upper black curve represents the experimental aging data collected at all temperatures used in this work, which represents the evolution of the relaxation time of samples without deformation at times $t = t_r + t_a$. The lower black line is the generic aging trajectory at $t_w/t_a$. This represents the aging trajectory of a sample that is freshly quenched at $t = t_a$ and defines the limit of rejuvenation. It is clear that the cycled glasses show lower relaxation times immediately after cycling compared to undeformed, aging glasses. Figure 8 also shows that the relaxation times for cycled samples return to the generic aging trajectory after recovery times close to the initial age $t_a$. This indicates that the data fall on path (b) in Figure 6 of ref [51], and that the glasses “remember” their initial age, rather than experience a full erasure of their thermal and mechanical history. Previous work[50] on similar lightly crosslinked PMMA glasses has also identified a range of recovery behavior after a reversing strain protocol that is consistent with the interpretations of ref [51]. Together, our new experiments and these previous experiments indicate that a considerable range of deformations at temperatures near $T_g$ do not give rise to overaging.

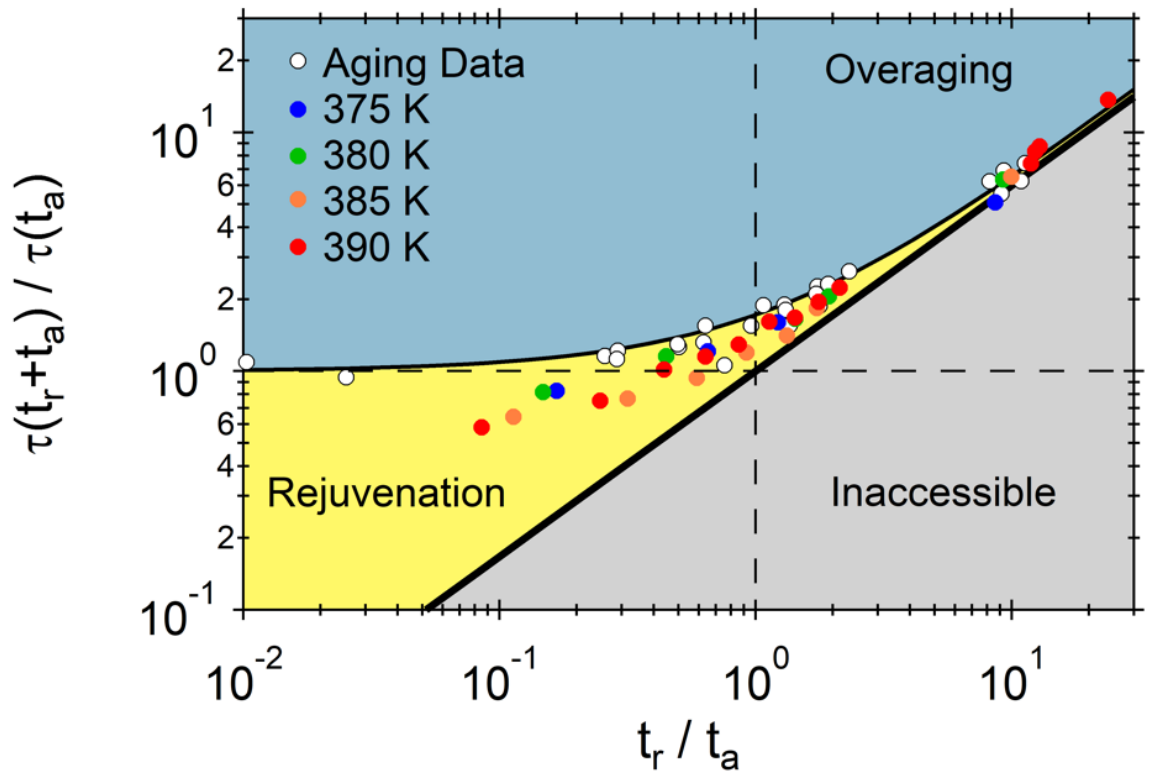

**Figure 8.** Recovery of segmental relaxation times of PMMA glasses after a single set of 5000 loading/unloading cycles. The vertical axis is a normalized relaxation time and the horizontal axis is a normalized recovery time. The diagonal line represents the generic aging trajectory at $t_w/t_a$. The observed relaxation times recover from a partially rejuvenated state back to the generic aging trajectory on a common time scale of $t_r \approx t_a$.

**Comparison to Overaging Simulations.** Overaging in computer simulations is typically identified as a deformation-induced decrease of the inherent structure energy that occurs more rapidly than is observed from physical aging alone. Though the inherent structure energy is not an experimental observable, previous simulation work on polymer glasses has established a correlation between inherent structure energy and molecular mobility.[52] Thus, it is reasonable to expect that the mobility of overaged glasses would be lower than a quiescently aged system. Even if molecular mobility is not perfectly correlated with inherent structure energy, the relaxation time is still directly relevant to material properties and controls the time scale for physical aging,[7] and it is at least as important as the inherent structure energy to identifying the effects of overaging.

Given these simulation results, why was lower mobility not observed as a result of loading/unloading in our experiments on PMMA glasses? One potentially important difference between simulations and the experiments performed here is the quench rate used to prepare the glasses. Simulations commonly quench on the order of 1 K/ps[25, 26] or faster,[20, 22] a quench rate nearly 14 orders of magnitude faster than our experimental cooling rate. Recent simulations have found that the overaging effect is highly dependent on the extent of annealing in the glass, such that more annealed glasses exhibit less or no overaging.[23, 27, 28] Given the difference in cooling rates, it is possible that we did not observe overaging in our experiments because the glasses were already reasonably annealed before the experiment began. Another related difference between simulations and our experiments is the deformation time relative to the glass preparation time. In simulation studies, the time spent cycling the glass is orders of magnitude longer than its undeformed age. This is not the case in the present experiments, where the deformation time and the preparation time are similar. Additionally, the present experiments are limited to deforming samples with an asymmetric loading/unloading protocol while recent overaging simulations have utilized symmetric cycling protocols. If the overaging response depends on the symmetry of the mechanical protocol, this would be important to understand.

**Comparison to Overaging Experiments.** In colloidal glasses, longer relaxation times induced by oscillatory shear compared to a purely aging system provide quite direct evidence for overaging.[29] For these experiments, the colloidal glass was subjected to a large shear to bring the sample to a reproducible, fully rejuvenated state. After the cessation of the initial large shear, the glass was allowed to age for 10 seconds and then subjected to 100 cycles of a smaller oscillatory shear deformation. It was observed that the relaxation time of the sheared glass was a factor of 2 slower than a glass that remained undeformed after the initial large shear. The authors concluded that dynamics in cycled glasses were slowed by the shear deformation. Though colloidal glasses and molecular glasses have distinct behavior,[53] the dynamics of colloidal glasses show some similarities to polymer glasses, such as exhibiting power-law aging behavior[29, 41, 42] as well as a power-law relationship between the relaxation time and deformation rate in the flow regime,[45, 46, 54] so one might expect to be able to observe overaging in the segmental dynamics of polymer glasses similar to that observed in the relaxation dynamics of colloidal glasses. However, our experiments did not induce overaging in PMMA glasses. This might be attributed to the difference in quenching protocols. The “mechanical quench” utilized to start the experiments on colloidal glasses is much faster than the thermal quench used to prepare the polymer glasses in this work. Similar to the interpretation of the simulation results, a faster quench rate or a shorter preparation time could be a potentially important factor to observe overaging. Alternatively, the nature of the mechanical quench may give rise the overaging result observed in colloidal glasses. Recent experiments on polystyrene glasses have identified additional relaxation processes after mechanical rejuvenation compared to a thermally rejuvenated sample.[55] The additional relaxation processes in a mechanically rejuvenated sample may effect a difference in the aging behavior compared to a thermally quenched sample. Initiating a cyclic loading/unloading experiment with a mechanical quench on a ductile polymer glass, rather than a thermal quench, would be an important extension of the present cyclic loading/unloading experiments.

Other experiments on polymeric[34, 35] and metallic glasses[30] have interpreted an increased yield stress compared to an undeformed, aging system as evidence of overaging. In experiments on fully amorphous polymer glasses, an increased yield stress can be induced by static or oscillating stress at temperatures ranging from $T_g$ – 70 K to $T_g$ – 145 K, a much lower temperature regime than is examined in the present work. The mechanical property measured in our experiments, the ratcheting compliance, shows typical aging behavior, but no signs of overaging. The difference in the aging behavior observed in our experiments and those of refs[34,35] might be due to the different temperature regimes. An alternate explanation might be that the ratcheting compliance is measured during cycling, while the yielding behavior is the result of a nonlinear deformation after the static or oscillating stress has been applied and removed. It is possible that the static or oscillating stress induces a change in the material that would not naturally occur during physical aging, such as orientation of polymer chains.[39] If this occurs, some or all of the enhanced yield stress might not result from overaging. Future experiments will replicate the experimental conditions of refs[34,35] and explore overaging in the segmental dynamics using the probe reorientation technique.

While the experimental conditions of the present work are relevant to polymers just below their glass transition temperatures, there is a vast parameter space beyond that explored here (colder temperatures, faster cooling rates, different modes of deformation, etc.). Clearly, our work does not generically rule out the existence of overaging in polymer glasses. However, in the parameter space that we are able to explore, our work finds no evidence for overaging.

**Comparison of Shift Factors for Different Observables.** Figure 9 shows the aging rate of lightly crosslinked PMMA glasses using the experimental techniques in the present work, as well as in the work of Ricci et al.[41] The aging rates measured with ratcheting compliance (RCC) and the standard creep compliance (SCC) data are determined from horizontal shift factors required for superposition while the aging rates measured from the probe reorientation technique (PR) and from linear stress relaxation tests (SR) are determined using segmental relaxation times extracted from fitting data with a KWW function. While not central to our current investigation, an interesting feature of Figure 9 is the difference in the aging rates among the measurement techniques. This is most clear at 385 K where the RCC and SCC data show $\mu \approx 1$, PR data shows $\mu \approx 0.8$ and SR data shows $\mu \approx 0.7$. We do not attribute these differences to random experimental error; for example, the probe reorientation results presented here and in reference 41 are in good agreement. Neither can one say that the results are explained by a difference between optical and mechanical experiments as the compliance and stress relaxation (both mechanical measurements) differ substantially from each other. It has been observed previously that measuring aging kinetics through different observables can lead to different results; for example, the time required for a polymer glass to reach equilibrium after a large (nonlinear) temperature jump can be different when monitoring the system's enthalpy, creep compliance, or volume.[56] Figure 9 might also be viewed this way. One possible explanation for this effect is that different properties depend on the underlying relaxation time distribution in different ways and that the shape of distribution is changing during aging. This provides one way to understand Figure 9 but further work would be required to test this view.

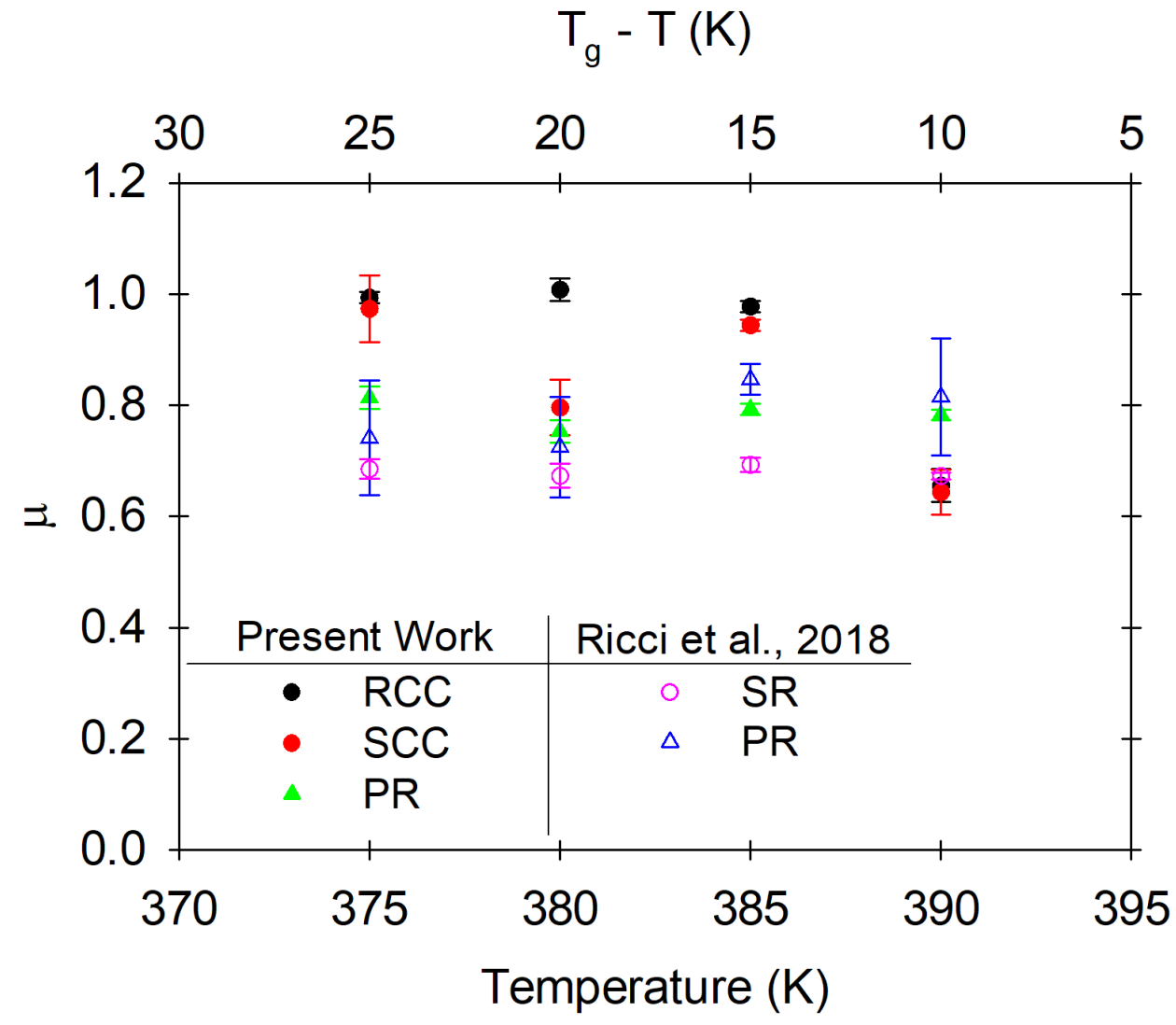

**Figure 9.** Aging rate of lightly crosslinked PMMA glasses measured using various techniques (RCC – ratcheting creep compliance; SCC – standard creep compliance; SR – stress relaxation; PR – probe reorientation). Filled data points are from the present work (error bars represent standard error) while the open data points are from ref[41] (error bars represent 90% confidence interval). Aging rates vary depending on the property being measured.

## Conclusion

In this work, we have shown that, for a selected set of conditions, the segmental dynamics of lightly crosslinked PMMA glasses do not become slowed or overaged by tensile loading/unloading cycles, but are rather accelerated or left unaffected, as measured by the probe reorientation technique. Similarly, the mechanical properties of lightly crosslinked PMMA glasses show no indications of overaging as judged by the ratcheting compliance of the material. When subjected to a single set of 5000 cycles, PMMA glasses show rejuvenation and subsequent recovery that are consistent with a generalized recovery path that indicates a memory of their initial age, as opposed to a full erasure of their thermal and mechanical history. Glasses subjected to multiple sets of 5000 cycles show clear incremental rejuvenation in their microscopic dynamics, but the evolution of their macroscopic ratcheting compliance is consistent with a linear deformation, indicating that the probe reorientation technique is more sensitive to nonlinear effects of deformation in polymer glasses.

Though this work does not explore the vast parameter space potentially relevant for overaging, it does examine conditions that are relevant to real polymer systems and, significantly, we find no indications of overaging. This suggests that slow-cooled polymer glasses just below their glass transition temperature are unlikely to be mechanically annealed to a more thermodynamically stable state, as can be done in simulated glasses.[21] This also potentially raises questions about constitutive models that use overaging as a key parameter to estimate the time-to-failure of a load-bearing polymer glass.[35, 40] More broadly, the results of this work provide further insight into how the segmental dynamics of polymer glasses are affected by cyclic loading/unloading, which is important for improved prediction of the long-term behavior of polymer glasses in engineering applications.

**Supporting Information.** Comparison between ratcheting compliance and standard creep compliance, vertical shift factors for time-aging time superposition of ratcheting compliance, superposition curves of ratcheting compliance with various $\varepsilon_{cycle}$

## Acknowledgements

We thank the National Science Foundation (DMR-1708248 and DMR-2002959) for support of this work.

**Supporting Information**

Rejuvenation versus overaging: The effect of cyclic loading/unloading on the segmental dynamics of PMMA glasses

Trevor Bennin*, Enran Xing, Josh Ricci, M. D. Ediger

Department of Chemistry, University of Wisconsin – Madison, Madison, Wisconsin 53706, United States

*Corresponding Author (tbennin@wisc.edu)

## Comparison between ratcheting compliance and standard creep compliance

The ratcheting compliance for PMMA glasses (as defined in the main text) is observed to be the same (after multiplication by a factor of 1.76) as the standard creep compliance measured using a small static stress (within the linear regime). This is shown for compliance curves collected at 375 K in Figure S1. The reason for this multiplicative factor is unclear. One possibility is related to the inhomogeneous thickness profile of the samples used; the stresses used to calculate the compliance were based on the dimensions of the sample at the thinnest region which will overestimate the average stress applied to the sample. However, we do not expect this to be a large enough effect to fully explain the factor of 1.76.

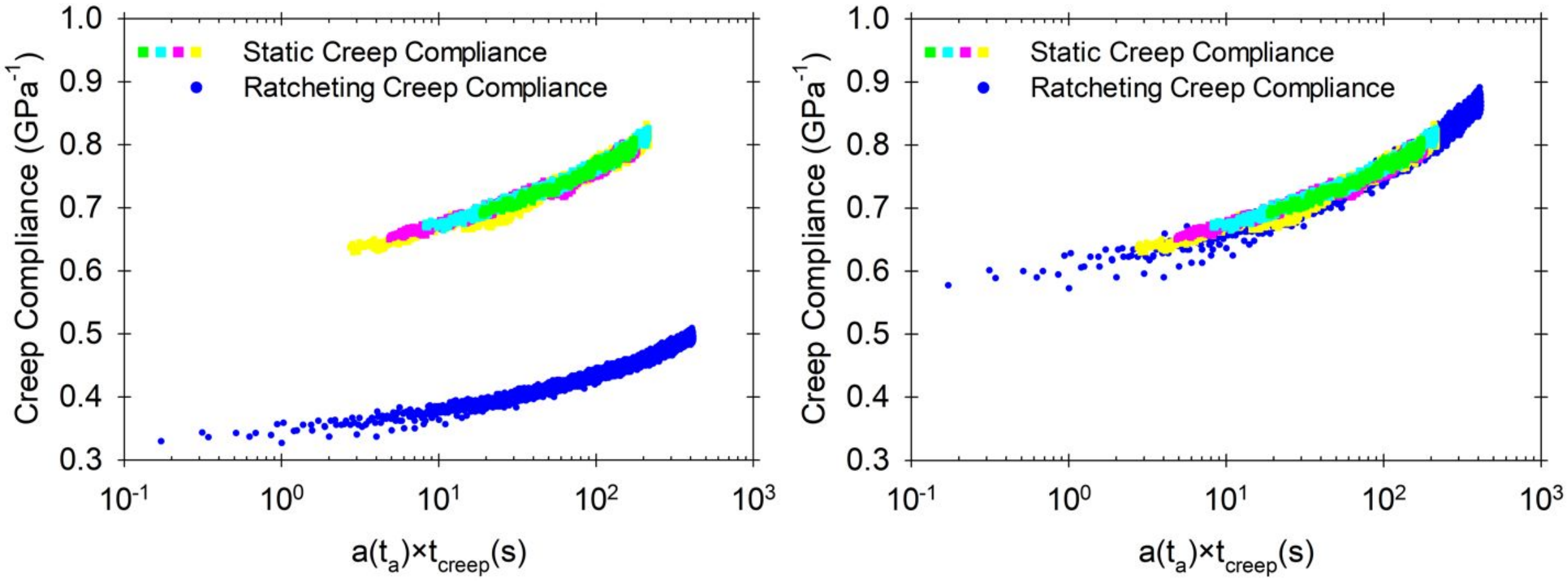

**Figure S1.** Comparison of the ratcheting compliance and the standard creep compliance at 375 K. The ratcheting compliance superposes with the static creep compliance when multiplied by a factor of 1.76.

The standard creep compliance data was collected on samples at temperatures ranging from 375 K to 390 K. The applied stress for each creep deformation was as follows: 3.65 MPa at 375 K, 3.34 MPa at 380 K, 3.13 MPa at 385 K, and 2.65 MPa at 390 K. The low applied stresses are expected to be within the linear regime. The standard creep compliance was measured using a protocol inspired by Struik: for any creep deformation, the creep time was limited to 10% of the initial aging time ($t_{creep}/t_a \leq 0.1$) and each subsequent creep deformation occurs at an aging time at least twice as large as the previous creep deformation ($t_{a,n+1} \geq 2 \times t_{a,n}$). These two conditions minimize the effects of aging and the previous deformation on any single creep compliance measurement. Physical aging of the standard creep compliance was analyzed using time-aging time superposition (as described in the main text), which provided horizontal shift factors $a(t_a)$. The aging rate, $\mu = -\frac{d\log(a(t_a))}{d\log(t_a)}$, for the ratcheting compliance and the standard creep compliance is observed to be the same as shown in Figure 9 of the main text.

**Time-aging time superposition of ratcheting compliance**

As described in the main text, the superposition procedure for the ratcheting compliance employed both horizontal and vertical shifts. The vertical shifts used for Figure 5 of the main text are shown in Figure S2, along with lines indicating the scatter in the data for the three low temperature data sets. The vertical shift factors are small (on the order of the noise in a compliance measurement) and not systematic with time for data collected at 375 K, 380 K and 385 K and are thus attributed to measurement error. Larger shifts that grow with aging time are required for the 390 K data. This is attributed to softening that occurred due to deformation; the average stress measured during cycling decreased with each subsequent set of cycles .

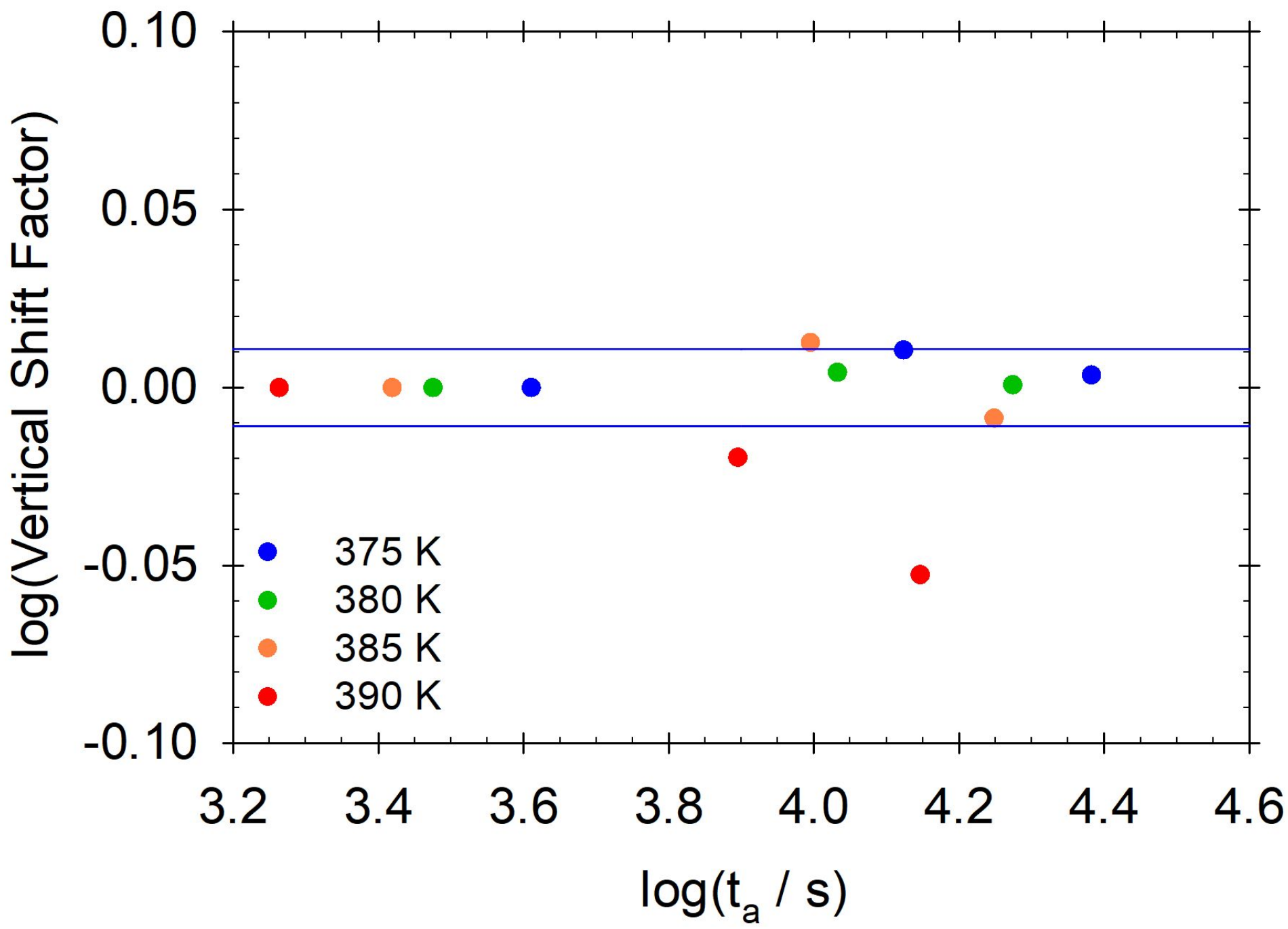

**Figure S2.** Vertical shift factors employed for time-aging time superposition of ratcheting compliance and an estimate of the scatter in the low temperature compliance data. The vertical shifts factors used for superposition at 375 K, 380 K and 385 K are on the order of the noise in the measurement.

In addition to the superimposed ratcheting compliance curves shown in Figure 5 of the main text, the ratcheting compliance data for glasses subjected to cyclic loading/unloading at various $\varepsilon_{cycle}$ can also be superimposed. The superposition of the ratcheting compliance curves for $\varepsilon_{cycle} = 0.003$ and $0.0072$ is shown in Figure S3.

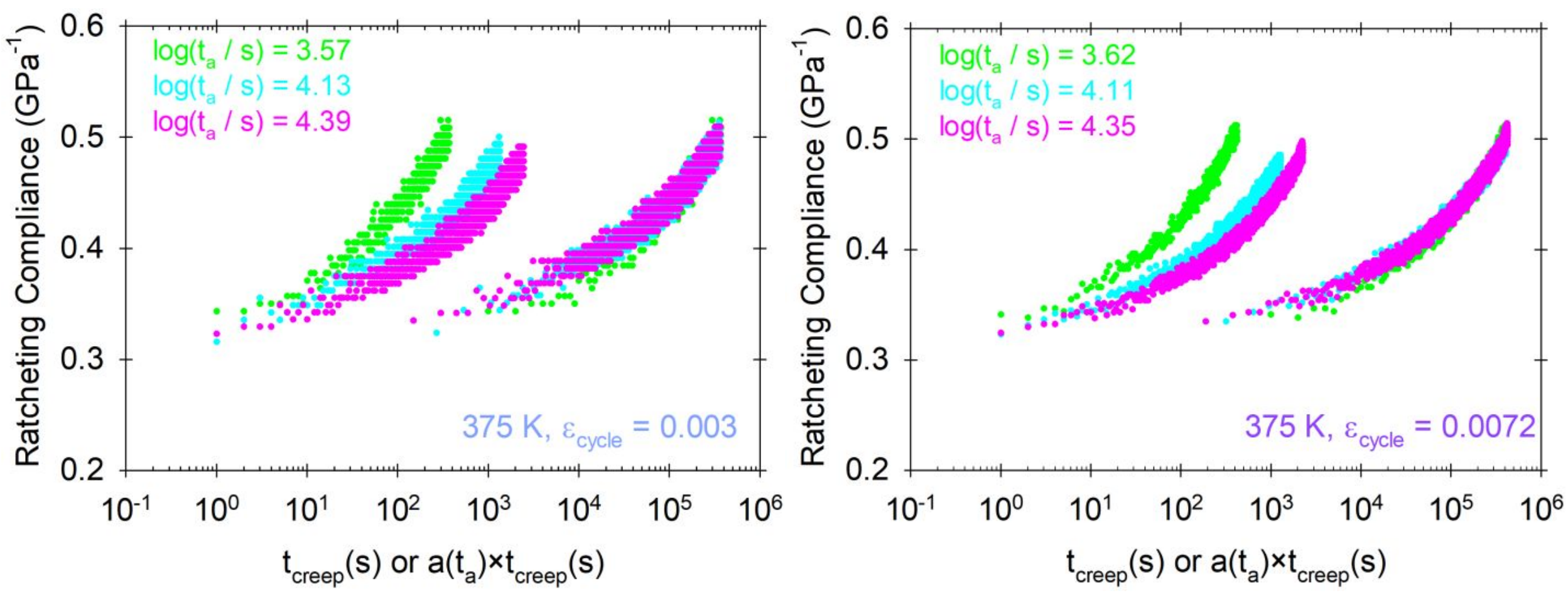

**Figure S3.** Time-Aging Time superposition for PMMA ratcheting compliance at 375 K and various $\varepsilon_{cycle}$.